\documentclass[aps,prl,twocolumn,showpacs]{revtex4-1}

\usepackage{graphicx}

\begin{document}

\title{How weak values emerge in joint measurements on cloned quantum systems}


\author{Holger F. Hofmann}
\email{hofmann@hiroshima-u.ac.jp}
\affiliation{
Graduate School of Advanced Sciences of Matter, Hiroshima University,
Kagamiyama 1-3-1, Higashi Hiroshima 739-8530, Japan}
\affiliation{JST, CREST, Sanbancho 5, Chiyoda-ku, Tokyo 102-0075, Japan
}

\begin{abstract}
A statistical analysis of optimal universal cloning shows that it is possible to identify an ideal (but non-positive) copying process that faithfully maps all properties of the original Hilbert space onto two separate quantum systems. The joint probabilities for non-commuting measurements on separate clones then correspond to the real parts of the complex joint probabilities observed in weak measurements on a single system, where the measurements on the two clones replace the corresponding sequence of weak measurement and post-selection. The imaginary parts of weak measurement statics can be obtained by replacing the cloning process with a partial swap operation. A controlled-swap operation combines both processes, making the complete weak measurement statistics accessible as a well-defined contribution to the joint probabilities of fully resolved projective measurements on the two output systems. 
\end{abstract}


\maketitle

Quantum mechanics cannot be explained by a simple realistic model of physical properties, because it is impossible to jointly determine the values of non-commuting observables. If a final measurement on an initial state $\mid \psi \rangle$ determines the values of an observable $\hat{A}$ with eigenstates $\mid a \rangle$, it is impossible to know what the result for an observable $\hat{B}$ with eigenstates $\mid b \rangle$ would have been for the same system, and vice versa. Specifically, a measurement of $\hat{B}$ performed after the measurement of $\hat{A}$ will  not have the same result as a measurement of $\hat{B}$ performed before the measurement of $\hat{A}$, since the projection of the quantum state onto an eigenstate of $\hat{A}$ completely changes the statistics of $\hat{B}$. 

In measurement theory, the projection of the state is related to decoherence in the interaction with the meter system. One way to avoid the changes in the statistics of $\hat{B}$ is therefore to weaken the measurement interaction until the decoherence effects are negligible. This approach to quantum measurement was introduced in 1988 by Aharonov, Albert and Vaidman, and is now widely applied to the study of quantum systems \cite{Aha88,Res04,Jor06,Wil08,Lun09,Yok09,Lun10,Gog11,Lun11,Koc11,Iin11}. In weak measurements, the value of an observable $\hat{A}$ between the preparation of an initial state $\mid \psi \rangle$ and a final measurement of $\mid b \rangle$ is obtained from the small shift of a meter caused by a weak interaction. The average result obtained from many measurements of $\mid \psi \rangle$ with the same final outcome $\mid b \rangle$ is given by the weak value,
\begin{equation}
\label{eq:weakvalue}
\frac{\langle b \mid \hat{A} \mid \psi \rangle}{\langle b \mid \psi \rangle}
= \sum_a A_a \frac{\langle b \mid a \rangle\langle a \mid \psi \rangle}{\langle b \mid \psi \rangle}.
\end{equation}
The right hand side of this equation shows the relation between the complex weak value and the eigenvalues $A_a$. The analogy with conventional quantum statistics suggests that the weak value is an average of the eigenvalues $A_a$ for a complex conditional probability of the eigenstate $\mid a \rangle$ in the pre- and post-selected ensemble defined by $\mid \psi \rangle$ and $\mid b \rangle$. This interpretation is also consistent with the notion that probabilities should be defined as expectation values of the projection operators $\mid a \rangle\langle a \mid$, since the complex conditional probabilities in Eq.(\ref{eq:weakvalue}) are given by the weak values of $\mid a \rangle\langle a \mid$. Significantly, the negative real parts of these weak conditional probabilities can explain quantum paradoxes, as demonstrated in a number of recent experiments \cite{Res04,Jor06,Wil08,Lun09,Yok09,Lun10,Gog11,Lun11,Koc11}. 

Based on the statistical interpretation of weak values, it is possible to define a joint probability of the measurement outcomes $\mid a \rangle$ and $\mid b \rangle$ \cite{Joh07b,Hof10,Hos10,Lun12,Hof12}. Specifically, the joint probability of $a$ and $b$ is obtained by multiplying the conditional probabilities of $a$ with the appropriate probability of $b$ given by $|\langle b \mid \psi \rangle|^2$ to obtain 
\begin{equation}
\label{eq:weakjoint}
\rho_{\mbox{weak}}(a,b) = \langle \psi \mid b \rangle\langle b \mid a \rangle\langle a \mid \psi \rangle. 
\end{equation}
Weak values therefore suggest that the correct joint probability of $a$ and $b$ is obtained by using the expectation value of the operator product of the projectors on $\mid a \rangle $ and  $\mid b \rangle$. Remarkably, this definition of complex joint probabilities was already introduced by Kirkwood in 1933 as an alternative to the Wigner function, and later generalized to discrete systems \cite{Kir33,Dir45,Pim88}. For arbitrary mixed states, the complex joint probability $\rho_{\mbox{weak}}(a,b)$ provides a complete characterization of the density matrix for any pair of basis sets $\{\mid a \rangle\}$ and $\{\mid b \rangle\}$ with non-zero overlaps between all $\mid a \rangle$ and $\mid b \rangle$ \cite{Joh07b,Lun12,Hof12}. 
Since weak values thus correspond to a fundamental expression of quantum statistics that was already studied long before weak measurements were introduced, one might expect that other implementations of joint measurements should reveal the same correlations between non-commuting observables. In this letter, it is shown that the joint probabilities derived from the statistical interpretation of weak measurements also appear in the correlations between cloned quantum systems, indicating that cloning can be used to confirm the statistical correlations between non-commuting observables described by weak values in a physical context that is completely different from weak measurements. 

Although cloning usually focuses on the attempt to copy the quantum state, the statistical interpretation of quantum mechanics implies that this can only be done by copying the actual physical properties, whether defined by the state or not. Ideal universal cloning should therefore copy all physical properties faithfully to both outputs, resulting in perfect correlations between the clones for all measurements. Of course, such an ideal cloning process is impossible, as stated by the well-known no-cloning theorem \cite{Sca05}. However, approximate cloning can be optimized by exploiting the symmetry of the quantum states \cite{Wer98}. In the following, it is shown that the errors in the optimal cloning process can be interpreted as an accidental distribution of white noise to one of the outputs, leaving a contribution that defines ideal cloning in terms of a non-positive linear map. Significantly, this ideal cloning process describes perfect correlations between all physical properties of the output systems, supporting the notion that the cloning process actually copies all physical properties equally, whether they are eigenvalues of the input state or not. 

Since the physical properties of the output clones in the ideal cloning term are perfectly correlated, separate measurements performed on the two clones can be interpreted as simultaneous measurements on the same system. The joint probabilities obtained from the clones therefore represent the correlations between non-commuting properties in the single input system before the cloning process. Significantly, the analysis of the ideal cloning term reproduces the joint probabilities previously observed in weak measurements, as given by Eq. $(\ref{eq:weakjoint})$. The only significant difference is that cloning does not define a temporal sequence, resulting in equal contributions from the sequence $(a,b)$ and $(b,a)$. Since these contributions have opposite imaginary parts, only the real part of the complex joint probability $\rho_{\mbox{weak}}(a,b)$ appears in the correlations between the clones. The imaginary parts can be obtained by a partial swap operation between the input state and white noise, which differs from the cloning process only in the phase of the superposition between swap and no swap. It is therefore possible to recover the temporal order of weak measurements from the direction of the partial swap operation that can be used to realize the optimal cloning process.


The starting point of the following analysis is the formulation of optimal quantum cloning as a map between the single $d$-dimensional Hilbert space of the input and the bosonic (or positive parity) states of the two output systems given by $(\mid \psi; m \rangle+\mid m; \psi \rangle)$ \cite{Sca05,Wer98}. By mixing over a complete orthogonal basis set $\{\mid m \rangle\}$, all information other than the input state $\mid \psi \rangle$ is eliminated. The output density matrix of the optimal cloning process can then be written in terms of the contributions from $\mid \psi \rangle$ in output 1, from $\mid \psi \rangle$ in output 2, and from the coherence between the two,
\begin{eqnarray}
\label{eq:clone}
\lefteqn{E_{\mathrm{clone}}(\mid \psi \rangle\langle \psi \mid) =}
\nonumber \\
&& \frac{1}{2(d+1)}\big(\mid \psi \rangle\langle \psi \mid \otimes \hat{I}
+ \hat{I}  \otimes  \mid \psi \rangle\langle \psi \mid
\nonumber \\ &+&
\sum_m \! \left( \mid \psi \rangle \langle m \mid \! \otimes \! \mid m \rangle \langle \psi \mid + \mid m \rangle \langle \psi \mid \! \otimes \! \mid \psi \rangle \langle m \mid \right)\, \big).
\end{eqnarray}
Here, $\{\mid m \rangle\}$ can be any orthogonal basis of the single system Hilbert space.
The first two terms can be interpreted as valid quantum operations in their own right, corresponding to the random distribution of the input state and a completely random state to the outputs with a total probability of $d/(d+1)$. The remaining output probability of $1/(d+1)$ is then provided by the coherence term, which describes a non-positive map of the input to the two output systems. 

In its most compact form, the contribution of the coherence term to the output can be represented by an operator $\hat{C}_{\psi}$ that is not self-adjoint and hence describes complex-valued contributions to the statistics,
\begin{equation}
\hat{C}_{\psi} =  \sum_m \mid \psi \rangle \langle m \mid \otimes \mid m \rangle \langle \psi \mid.
\end{equation} 
By itself, this operator defines a non-positive linear map of the input state $\mid \psi \rangle\langle \psi \mid$ that faithfully copies the input state to both outputs. Specifically, the local states obtained when tracing out one of the two output systems are both given by $\mid \psi \rangle$,
\begin{equation}
\mbox{Tr}_1 \left(\hat{C}_{\psi}\right)=
\mbox{Tr}_2 \left(\hat{C}_{\psi}\right)= \mid \psi \rangle 
\langle \psi \mid.
\end{equation}
Thus the non-positive state $\hat{C}_{\psi}$ has a perfect cloning fidelity of one. Moreover, this state also describes perfect correlations between all quantum fluctuations of $\mid \psi \rangle$. If the same projective measurement onto an orthogonal basis $\{\mid a \rangle\}$ is performed in both systems, $\hat{C}_{\psi}$ only contributes to the probabilities of finding the same result in both outputs, as shown by the projection
\begin{equation}
\langle a; a^\prime \mid \hat{C}_{\psi}\mid a; a^\prime \rangle =
\delta_{a,a^\prime} |\langle a \mid \psi \rangle|^2.
\end{equation}
This is the signature of a physical copy of the input system: even an unknown property $a$ is faithfully duplicated. It may therefore be justified and useful to think of the quantum coherence $\hat{C}_{\psi}$ in the optimal cloning process as a duplication of all physical properties in Hilbert space, where the quantum coherences in the initial Hilbert space are transferred to directly observable correlations between the separate Hilbert spaces of two systems. 


If different measurements are performed in the two output systems, the contributions of $\hat{C}_{\psi}$ to the joint probabilities of ${\mid a \rangle}$ and ${\mid b \rangle}$ are given by 
\begin{eqnarray}
\label{eq:comprob}
\rho_C(a,b) &=&
\langle a;b \mid \hat{C}_{\psi} \mid a; b \rangle
\nonumber \\
&=&  \langle \psi \mid b \rangle \langle b \mid a \rangle \langle a \mid \psi \rangle.
\end{eqnarray}
As explained in the introduction, this is exactly equal to the complex joint probability $\rho_{\mbox{weak}}(a,b)$ in Eq.(\ref{eq:weakjoint}), which is obtained from a weak measurement of $\mid a \rangle\langle a \mid$ followed by a final measurement of $\mid b \rangle\langle b \mid$. Specifically, the weak values of $\mid a \rangle\langle a \mid$ followed by a post-selection of $\mid b \rangle$ correspond to the conditional probabilities obtained by dividing $\rho_C(a,b)$ by the total probability of $b$ given by $|\langle b \mid \psi \rangle|^2$. The weak value of $\hat{A}$ can then be expressed as a conditional average of the joint probability $\rho_C(a,b)$,
\begin{equation}
\frac{\langle b \mid \hat{A} \mid \psi \rangle}{\langle b \mid \psi \rangle}
= \sum_{a} A_a \; \frac{\rho_C(a,b)}{\sum_{a^\prime} \rho_C(a^\prime,b)}.
\end{equation}
The statistical interpretation of weak values is therefore consistent with the correlations independently obtained from quantum cloning. 


It might seem a bit strange that the ideal cloning process $\hat{C}_\psi$ corresponds to a specific sequence of $a$ and $b$. In the real cloning process given by Eq.(\ref{eq:clone}), there is no preferred sequence since the output includes an equal mixture of the ideal cloning processes $\hat{C}_\psi$ and $\hat{C}_\psi^\dagger$ corresponding to the sequences $(a,b)$ and $(b,a)$, respectively. As a consequence, the imaginary parts of the joint probabilities $\rho_C(a,b)$ cancel, and only the real parts show up in the experimentally accessible joint probabilities obtained from an optimal cloning process. Specifically, the actual output probabilities $p(a,b)$ measured in the output of the optimal cloning process are given by
\begin{eqnarray}
p(a,b) &=& \langle a;b \mid E_{\mbox{clone}}(\mid \psi \rangle\langle \psi \mid) \mid a; b \rangle
\nonumber
\\
&& \hspace{-1.4cm} = \frac{1}{2(d+1)}\left(
|\langle a \mid \psi \rangle|^2\!
+ |\langle b \mid \psi \rangle|^2\!
+ 2 \mathrm{Re} \left(\rho_C(a,b)\right)
\right).
\end{eqnarray}
In addition to the real part of the complex joint probability $\rho_C(a,b)$, cloning errors result in a ``background'' obtained from the probabilities $|\langle a \mid \psi \rangle|^2$ and $|\langle b \mid \psi \rangle|^2$. Intuitively, this corresponds to the random results for $a$ or $b$ obtained when the cloning process distributed the white noise input to one of the outputs. Since the marginal probabilities $p(a)$ and $p(b)$ obtained for each system can be used to determine $|\langle a \mid \psi \rangle|^2$ and $|\langle b \mid \psi \rangle|^2$, it is a straightforward matter to subtract the background noise and to determine the real part of $\rho_C(a,b)$ from $p(a,b)$.


For completeness, it would also be desirable to obtain the imaginary part of the cloning statistics. In weak measurements, the change from real to imaginary weak values is achieved by replacing the weak measurement operator with a weak unitary transformation generated by the projector on $\mid a \rangle$ \cite{Lun11,Joz07,Hof11b,Dre12,Lun12}. Effectively, this corresponds to a phase change in the superposition between no measurement and projective measurement that characterizes the weak measurement interaction. In the present case, a similar effect can be achieved by changing the phase in the superposition between the distribution of the initial state to system 1 and the distribution to system 2. In terms of quantum logic operations, the distribution of the input can be achieved by performing or not performing a swap operation. The eigenstates of this operation are the states with positive (bosonic) and negative (fermionic) parity under the swap operation. The imaginary superposition of swap and no swap results in a unitary operation with eigenvalues of $(1+i)/\sqrt{2}$ and $(1-i)/\sqrt{2}$ for the positive and negative parity states. The operation thus corresponds to a partial swap represented by the square root of the unitary operator for the complete swap. The positive linear map of this partial swap or root-swap operation on an input product state of $\mid \psi \rangle$ and the maximally mixed state $\hat{I}/d$ is given by
\begin{eqnarray}
\lefteqn{E_{\sqrt{\mathrm{swap}}}(\mid \psi \rangle\langle \psi \mid)=}
\nonumber \\ &&
\frac{1}{2 d}\big(\mid \psi \rangle\langle \psi \mid \otimes \hat{I}
+ \hat{I} \otimes \mid \psi \rangle\langle \psi \mid 
- i \left(\hat{C}_\psi - \hat{C}_\psi^\dagger \right)\big).
\end{eqnarray}
Thus the imaginary part of the expectation values of $\hat{C}_\psi$ given by $\rho_C(a,b)$ represent the (real) correlations that build up between the two systems during the continuous transfer of quantum information from one system to the other in a unitary swap operation. Essentially, the correlations of the partial swap operations provide a complete map of the dynamical structure of Hilbert space that complements the statistical structure observed in the correlations between clones \cite{Hof11b}. 


Finally, it might be worth noting that an efficient combination of optimal cloning and partial swaps can be realized by a controlled-swap operation with a control qubit input in a superposition state of $\mid 0_z \rangle+\mid 1_z \rangle$. Any superposition of swap and no swap can then be accessed by simply measuring the corresponding superposition in the control qubit output. Since $\hat{C}_\psi$ is a direct representation of the coherence between swap and no swap, its contributions to the joint probabilities of $a$ and $b$ can be obtained from the differences between the joint probabilities obtained for opposite control qubit coherences. Specifically, the real part of $\rho_C(a,b)$ can be obtained from the difference in the joint probabilities for the positive and negative superpositions of 
$\mid 0_z \rangle$ and $\mid 1_z \rangle$ given by the $X$-basis eigenstates $\mid 0_x \rangle$ and $\mid 1_x \rangle$, while the imaginary part is given by the differences between the corresponding $Y$-basis eigenstates $\mid 0_y \rangle$ and $\mid 1_y \rangle$. The quantum controlled-swap then permits complete quantum tomography of the input state based on two non-commuting von Neumann measurements performed on the separate output systems and two different measurements on the control qubit output,
\begin{eqnarray}
\mathrm{Re}(\rho_C(a,b))&=& d \left(p(0_x,a,b)-p(1_x,a,b)\right)
\nonumber \\
\mathrm{Im}(\rho_C(a,b))&=& d \left(p(0_y,a,b)-p(1_y,a,b)\right).
\end{eqnarray}
The complete density matrix of an unknown input state $\hat{\rho}$ can then be reconstructed directly by
\begin{equation}
\hat{\rho}=\sum_{a,b} \rho_C(a,b) \frac{\mid a  \rangle \langle b \mid}{\langle b \mid a \rangle}.
\end{equation}
The quantum controlled-swap thus maps the complete quantum coherence of the input onto the correlations between $a$ and $b$ in the output. In contrast to weak measurements, no approximate limit is required and the statistical structure of Hilbert space appears as a well-defined part of conventional measurement statistics. In the controlled-swap, the subtraction of background noise is achieved by taking the difference between the probabilities of opposite measurement outcomes for the control qubit, so complex probabilities are observed directly as polarizations of the qubit in the $XY$-plane of the Bloch sphere. Note that this procedure is technically similar to the observation of the complex wave function in the polarization of the meter photon reported in \cite{Lun11}. However, the controlled-swap achieves much higher visibilities, since the measurements are not performed in the weak interaction limit.


The results presented above indicate that the complex joint probabilities obtained by multiplying the non-commuting measurement operators provide a consistent representation of statistical correlations between non-commuting measurements in quantum mechanics. In particular, the non-positive linear map represented by $\hat{C}_\psi$ appears to be the quantum mechanical analog of a faithful copy of the input system to both output systems. Since this linear map works for all input states and does not depend on the choice of measurements performed on the output, the origin of the correlations between the measurement outcomes can be traced back to the single Hilbert space of the input state. Thus, quantum cloning and partial swaps provide an approach to joint measurements that avoids many of the ambiguities of sequential or joint measurements performed on the same system. 

It is especially remarkable that the results correspond to those observed in weak measurement, since the physics of the measurement procedure are obviously quite different. While weak measurements distinguish between the weakly measured result $a$ and the post-selected result $b$, cloning and partial swaps are completely symmetric in the output measurements. The measurement interaction is well separated from the cloning process, so the correlations observed cannot be induced by the dynamics of the measurement interaction. The observation of weak measurement statistics in cloning and in partial swaps therefore supports the interpretation of weak measurements as back-action free measurements of joint probabilities \cite{Hof10,Hos10,Lun12,Hof12}. 

In conclusion, the effects of quantum coherence in optimal cloning processes can be described by a non-positive linear map that represents the quantum mechanical analog of a perfect copying process. This perfect copying process maps the statistical correlations of non-commuting measurement in a single Hilbert space onto two separate systems, where both measurements can be performed jointly. The statistics obtained from the joint measurements correspond to those obtained in weak measurements and thereby support the interpretation of weak measurements in terms of joint and conditional probabilities. A statistical analysis of quantum mechanics in terms of complex joint probabilities and non-positive maps can thus help to uncover unifying principles in quantum information and improve our understanding of the physics behind non-classical effects.   


Part of this work has been supported by the Grant-in-Aid program of the Japan Society for the Promotion of Science, JSPS.


\vspace{0.5cm}

\begin{thebibliography}{xyz00}


\bibitem{Aha88}
Y. Aharonov, D. Z. Albert, and L. Vaidman, Phys. Rev. Lett. {\bf 60}, 1351 (1988).


\bibitem{Res04}
K. J. Resch, J. S. Lundeen and A. M. Steinberg, Phys. Lett. A {\bf 324}, 125 (2004).

\bibitem{Jor06}
A. N. Jordan, A. N. Korotkov, and M. B\"uttiker, Phys. Rev. Lett. {\bf 97}, 026805 (2006).

\bibitem{Wil08}
N. S. Williams and A. N. Jordan, 
Phys. Rev. Lett. {\bf 100}, 026804 (2008).

\bibitem{Lun09}
J. S. Lundeen and A. M. Steinberg, Phys. Rev. Lett. {\bf 102}, 020404 (2009).

\bibitem{Yok09}
K. Yokota, T. Yamamoto, M. Koashi, and N. Imoto, New J. Phys. {\bf 11}, 033011 (2009).

\bibitem{Lun10}
A. P. Lund and H. M. Wiseman, New J. Phys. {\bf 12}, 093011 (2010).

\bibitem{Gog11}
M. E. Goggin, M. P.Almeida, M. Barbieri, B. P. Lanyon, J. L. O'Brien, A. G. White, and G. J. Pryde, Proc. Natl. Acad. Sci. U. S. A. {\bf 108}, 1256 (2011).

\bibitem{Lun11}
J. S. Lundeen, B. Sutherland, A. Patel, C. Stewart, and C. Bamber, Nature (London) {\bf 474}, 188 (2011).

\bibitem{Koc11}
S. Kocsis, B. Braverman, S. Ravets, M. J. Stevens, R. P. Mirin, L. K. Shalm, and  A. M. Steinberg,
Science {\bf 332}, 1170 (2011).

\bibitem{Iin11}
M. Iinuma, Y. Suzuki, G. Taguchi, Y. Kadoya, and H. F. Hofmann, New J. Phys. {\bf 13}, 033041 (2011). 


\bibitem{Joh07b}
L. M. Johansen, Phys. Rev. A {\bf 76}, 012119 (2007).

\bibitem{Hof10}
H. F. Hofmann, Phys. Rev. A {\bf 81}, 012103 (2010).  

\bibitem{Hos10}
A. Hosoya and Y. Shikano, 
J. Phys. A: Math. Theor. {\bf 43}, 025304 (2010).

\bibitem{Lun12}
J. S. Lundeen and C. Bamber, Phys. Rev. Lett. {\bf 108}, 070402 (2012). 

\bibitem{Hof12}
H. F. Hofmann, New J. Phys. {\bf 14}, 043031 (2012).


\bibitem{Kir33}
J. G. Kirkwood, Phys. Rev. {\bf 44}, 31 (1933).

\bibitem{Dir45}
P. A. M. Dirac, Rev. Mod. Phys. {\bf 17}, 195 (1945).

\bibitem{Pim88}
A. Pimpale and M. Razavy, Phys. Rev. A {\bf 38}, 6046 (1988).


\bibitem{Sca05}
For a review of quantum cloning, see
V. Scarani, S. Iblisdir, N. Gisin, and A. Acin, Rev. Mod. Phys. {\bf 77}, 1225 (2005).

\bibitem{Wer98}
R.F. Werner, Phys. Rev. A {\bf 58}, 1827 (1998).

\bibitem{Joz07}
R. Jozsa, Phys. Rev. A {\bf 76}, 044103 (2007).

\bibitem{Hof11b}
H. F. Hofmann, New J. Phys. {\bf 13}, 103009 (2011).

\bibitem{Dre12}
J. Dressel and A. N. Jordan, Phys. Rev. A {\bf 85}, 012107 (2012).

\end{thebibliography}
\end{document}